# Insights into teaching quantum mechanics in secondary and lower undergraduate education

K. Krijtenburg-Lewerissa,[1] H. J. Pol,[1] A. Brinkman,[2] and W. R. van Joolingen[3]

[1]*ELAN Institute for Teacher Training, University of Twente, 7500 AE Enschede, Netherlands*
[2]*MESA+ Institute for Nanotechnology, University of Twente, 7500 AE Enschede, Netherlands*
[3]*Freudenthal Institute for Science and Mathematics Education,
University of Utrecht, 3508 AD Utrecht, Netherlands*

This study presents a review of the current state of research on teaching quantum mechanics in secondary and lower undergraduate education. A conceptual approach to quantum mechanics is being implemented in more and more introductory physics courses around the world. Because of the differences between the conceptual nature of quantum mechanics and classical physics, research on misconceptions, testing, and teaching strategies for introductory quantum mechanics is needed. For this review, 74 articles were selected and analyzed for the misconceptions, research tools, teaching strategies, and multimedia applications investigated. Outcomes were categorized according to their contribution to the various subtopics of quantum mechanics. Analysis shows that students have difficulty relating quantum physics to physical reality. It also shows that the teaching of complex quantum behavior, such as time dependence, superposition, and the measurement problem, has barely been investigated for the secondary and lower undergraduate level. At the secondary school level, this article shows a need to investigate student difficulties concerning wave functions and potential wells. Investigation of research tools shows the necessity for the development of assessment tools for secondary and lower undergraduate education, which cover all major topics and are suitable for statistical analysis. Furthermore, this article shows the existence of very diverse ideas concerning teaching strategies for quantum mechanics and a lack of research into which strategies promote understanding. This article underlines the need for more empirical research into student difficulties, teaching strategies, activities, and research tools intended for a conceptual approach for quantum mechanics.

DOI: 10.1103/PhysRevPhysEducRes.13.010109

## I. INTRODUCTION

Quantum mechanics has gained a strong position in physics research and its applications. Developments in medical imaging, nanoscience, laser physics, and semiconductors are all based on quantum phenomena. Moreover, quantum mechanics is the foundation of completely new and promising technologies: quantum computers, quantum encryption, and quantum entanglement. Quantum mechanics has been an important part of university physics and engineering education for a long time, but the often abstract and mathematical teaching practices used have been in dispute for several years [1]. Currently, more emphasis is placed upon visualization and conceptual understanding [2,3]. This conceptual approach to quantum mechanics has made it possible to introduce quantum mechanics at an earlier stage, and therefore it has become part of the secondary school curriculum in many countries. Quantum mechanics has been part of the upper secondary school curriculum in England [4], Germany [5], Italy [6], and the USA [7] for several years. More recently, quantum mechanics has been incorporated in the Dutch [8] and the French [9] secondary school curricula, and in Norway new teaching modules have been designed and tested in the ReleQuant project [10].

Because quantum mechanics led to fundamental changes in the way the physical world is understood and how physical reality is perceived [11], quantum mechanics education is faced with several challenges. For instance, the introduction of probability, uncertainty, and superposition, which are essential for understanding quantum mechanics, is highly nontrivial. These concepts are counterintuitive and conflict with the classical world view that is familiar to most students. A radical change in thinking is needed [12] and ways to instigate conceptual change [13,14] should be investigated.

Several initiatives have been taken to improve students' understanding of quantum mechanics and resolve problems encountered in teaching quantum mechanics, including a review of misconceptions of upper level undergraduate students [15]. This review by Singh and Marshman gives a good overview of students' difficulties on an abstract and mathematical level. Introductory quantum mechanics courses mainly focus on the introduction of the main concepts and students' conceptual understanding hereof. Therefore, we reviewed articles covering educational research on quantum mechanics for the secondary and lower undergraduate level, aiming to answer the following question:

What is the current state of research on students' understanding, teaching strategies, and assessment methods for the main concepts of quantum mechanics, aimed at secondary and lower undergraduate education?

More specifically, we researched the following questions:
  (i) What learning difficulties do secondary and lower undergraduate level students encounter while being taught quantum mechanics?
  (ii) What instruments have been designed and evaluated to probe students' understanding on a conceptual level?
  (iii) What teaching strategies aimed at the secondary and lower undergraduate level have been tested, implemented, and evaluated for their influence on students' understanding?

The overview presented in this article therefore comprises (i) students' misconceptions and difficulties, (ii) research-based tools to analyze student understanding, and (iii) assessed instructional strategies, activities, and multimedia applications that improve student understanding.

## II. METHOD

For this study three databases were searched: Scopus, Web of Science, and ERIC. The following query was used to find appropriate articles, published in journals: "(quantum OR "de Broglie" OR "photoelectric effect") AND (student OR instruction) AND (concept OR understanding OR reasoning OR difficulties)." This search resulted in 471 articles from ERIC, Web of Science, and Scopus, published between 1997 and the present.

Subsequently, the results were filtered using the following criteria: (1) The article addresses the understanding of quantum concepts for secondary or undergraduate students in an educational setting, (2) the article includes an implementation and evaluation of its impact on understanding, (3) the article does not expect students to be familiar with mathematical formalism (e.g., Dirac notation, Hamiltonians, or complex integrals), and (4) the article mainly emphasizes physical aspects.

A total of 74 articles matched these criteria. These articles were analyzed for detected student difficulties, used research-based tools which measure student understanding, and assessed instructional strategies, multimedia applications, and activities. The following sections present these difficulties, tools, and teaching approaches, all categorized and analyzed for content, research methods, and value for teaching quantum mechanics in secondary and lower undergraduate education. Where needed, additional literature has been used to clarify or evaluate the findings in the selected literature.

## III. LEARNING DIFFICULTIES

For the development of effective teaching strategies, it is important to know what difficulties students have with quantum mechanics. Therefore this section gives an overview of findings for the first subquestion: "What learning difficulties do secondary and lower undergraduate level students encounter while being taught quantum mechanics?" To answer this question, the selected articles were all scanned for misconceptions concerning the topics shown in Table I. These topics were based on (1) the learning goals formulated by McKagan *et al.* [16], which were based on interviews with faculty members who had recently taught modern physics; and (2) learning goals determined in a Delphi study among Dutch experts in quantum mechanics [17], a method which uses consecutive questionnaires to explore consensus among experts [18]. The topics in Table I encapsulate the main topics found in introductory quantum mechanics curricula around the world [4–10]. This section gives an overview of misconceptions and learning difficulties found in the reviewed articles, organized by the topics in Table I. See the Appendix for more information concerning the research methods for articles discussed in this section.

### A. Wave-particle duality

The fact that tiny entities show both particle and wave behavior is called wave-particle duality. This phenomenon is in conflict with prior, classical reasoning. Several selected articles addressed the understanding of wave-particle duality [1,4,5,16,19–34]. Ireson and Ayene *et al.* researched existing student views of undergraduate students using cluster analysis [20,24,25]. Three clusters emerged: (1) Classical description, in which students describe quantum objects exclusively as particles or waves; (2) mixed description, in which students see that wave and particle behavior coexist, but still describe single quantum objects in classical terms; and (3) quasiquantum description, in which students understand that quantum objects can behave as both particles and waves, but still have difficulty describing events in a nondeterministic way. Similar categories of understanding were found by Greca and Freire [22] and Mannila et al. [26]. These clusters all depend on the extent to which students hold on to classical thinking and constitute a spectrum from misplaced classical thinking to correct quantum thinking. Table II gives an overview of misconceptions and learning difficulties encountered in the reviewed research, divided into these three clusters. In the following sections, the listed misconceptions are discussed in more detail.

TABLE I. Quantum topics used for the analysis of the selected articles.

| Wave-particle duality | Wave function | Atoms | Complex quantum behavior |
| --- | --- | --- | --- |
| Dual behavior of photons and electrons | Wave functions and potentials | Quantization and energy levels | Time dependent Schrödinger equation |
| Double slit experiment | Probability | Atomic models | Quantum states |
| Uncertainty principle | Tunneling | Pauli principle and spin | Superposition |
| Photoelectric effect | | | Measurement |

TABLE II. Misconceptions about wave-particle duality organized into three categories ranging from classical to quantum thinking.

|  | Classical description | Mixed description | Quasiquantum description |
|---|---|---|---|
| Photons or electrons | Electrons or photons are depicted as classical particles [1,4,5,16,20,22–25] | Electrons and photons follow a definite sinusoidal path [16,29,30] | Electrons are smeared clouds of charge [5,24,25] |
|  | Electrons or photons have definite trajectories [1,4,5,16,20,22–25] | Electrons are either a particle or a wave depending on other factors [21,29] | Electrons or photons are waves and particles simultaneously [20,30] |
|  | Light always behaves like a wave [24,25] | Equations of properties of light also apply to electrons [21] |  |
| Double slit experiment | Light has no momentum [1] | There is no relation between momentum and de Broglie wavelength [21,34] | There is no relation between momentum and interference pattern [21,34] |
|  | Photons and electrons deflect at a slit and subsequently move in a straight line [21] | No interference pattern appears with single photons and electrons [24–26] |  |
| Uncertainty principle | Uncertainty is due to external effects, measurement errors or measurement disturbance [5,20,32] |  |  |
| Photoelectric effect | Energy is transmitted by wave fronts, more wave fronts cause more energy [30] | Light collides with electrons [19,28] |  |
|  | The intensity of light influences the energy transferred to a single electron [27,28] |  |  |

### 1. Photons and electrons

In many cases electrons display particle properties, but that is not the entire picture. Electrons also exhibit wave properties, such as diffraction and interference. Conversely, light shows wave and particle behavior. Light diffracts, refracts, and shows interference, but additionally its energy is quantized, i.e., transferred in "packages." The reviewed literature showed that students have a range of different visualizations of photons and electrons, and many have difficulty juxtaposing wave and particle behavior. Research showed that many secondary and undergraduate students erroneously see electrons exclusively as particles and photons as bright spherical balls with a definite location or trajectory [4,5,22–25,29].

The wavelike behavior of electrons is hard to define, for electrons appear as bright spots on fluorescent screens in most of the textbook experiments. The wavelike behavior of electrons only appears in the distribution of these bright spots. Quantum mechanics does not describe an electron's path, only the probability of finding it at a certain location. Müller and Wiesner [5] observed that students sometimes falsely considered this wave behavior to be a cloud of smeared charge. McKagan et al. [16] and Olsen [29] reported that several secondary and undergraduate students considered the wave behavior of electrons to be a pilot wave, which forces the electron into a sinusoidal path.

Photons are also sometimes considered to move along sinusoidal paths [30], but Olsen observed that students showed less difficulty assigning both wave and particle behavior to light than to electrons [29]. Sen [31] observed that most students had a more scientific way of describing photons than electrons and ascribed this to the fact that photons are introduced later in the curriculum, which he believes to result in fewer misconceptions of photons at the start of undergraduate education.

### 2. Double slit experiment

The double slit experiment is used to illustrate the wavelike behavior of photons, electrons, buckyballs, and other small objects. These objects pass through a double slit, fall onto a detection screen, and cause an interference pattern. For electrons, this interference pattern appears only in the distribution of the bright spots. Understanding of the double slit experiment depends in part on the students' understanding of the wave and particle behavior of quantum objects. If students see photons as classical particles with definite trajectories, this influences their comprehension of this experiment. This can be seen by the fact that some secondary students considered photons to deflect at the slit edges and move in straight lines towards the screen [21]. Another common problem depends on incomplete understanding of the de Broglie wavelength. Students do not always understand the influence of velocity and mass on wavelength and the influence of wavelength on the interference pattern [21,34].

### 3. Uncertainty principle

The uncertainty principle states that there are certain properties that cannot simultaneously be well defined. An example thereof is the relation between position and momentum, for which the uncertainty principle is described as $\Delta x \Delta p \geq h/4\pi$. This equation shows that when one of the properties is determined with high precision, the outcome of a measurement of the other property becomes less certain. The uncertainty principle for position and momentum can intuitively be related to the wave behavior of small entities. For example, a strongly localized wave package is a superposition of many waves with varying wavelength and momentum. Ayene *et al.* [20] observed four categories of depictions of the Heisenberg uncertainty principle: (i) Uncertainty is erroneously described as a measurement error due to external effects, (ii) uncertainty is wrongly described as a measurement error due to error of the instrument, (iii) uncertainty is falsely thought to be caused by measurement disturbance, and (iv) uncertainty is correctly seen as an intrinsic property of quantum systems. Only a small number of students had views that fell within the fourth, correct, category. Müller and Wiesner [5] and Singh [32] also observed that secondary and undergraduate students attributed uncertainty to external effects. They reported that some students stated that uncertainty is caused by the high velocity of quantum particles.

### 4. Photoelectric effect

The photoelectric effect is the phenomenon by which materials can emit electrons when irradiated by light of sufficiently high frequency. This effect is used to show the particlelike behavior of light. This particlelike behavior emerges from the observation that the energy of the emitted electron depends solely on the frequency of the incident light, whereas the intensity of the light determines only the number of emitted electrons. For this subject Asikainen and Hirvonen [19] observed that some students confused the photoelectric effect with ionization. Their research also showed that certain students had difficulty with fully understanding how light and electrons interact, and how various aspects (work function, kinetic energy, cutoff frequency, and material properties) together constitute the photoelectric effect. McKagan *et al.* [27] observed that some undergraduate students could not distinguish between intensity and frequency of light, were unable to explain why photons are related to the photoelectric effect, falsely believed that an increase of light intensity will increase the energy transferred to a single electron, or incorrectly believed that a voltage is needed for the photoelectric effect. This last incorrect believe was also observed with secondary school students by Sokolowski [33]. Özcan [30] observed that undergraduate students' different models of light influenced their understanding of the photoelectric effect. Students who used the wave model falsely described the energy transfer in terms of vibrations, which were caused by wave fronts striking the metal. These students believed an increase in light intensity would lead to an increase in the number of wave fronts. Oh [28] observed that some undergraduate students wrongly thought that light reacts chemically with an electron, and others falsely believed that the intensity of light could influence if electrons were ejected or not.

### B. Wave functions

In this section the observed misconceptions concerning wave functions, potential wells, tunneling, and probability found in the selected articles [35–44] are presented. Articles matching our search criteria, which addressed the understanding of wave functions, described difficulties of undergraduate students only.

### 1. Wave functions and potential wells

Wave functions represent the state of particles. The wave function $\psi$ is not a physical wave, but a mathematical construct, which, for a bound electron, is specified by four quantum numbers, $n$, $l$, $m$ and $s$. $\psi$ contains all information of a system and predicts how particles will behave given a specific potential. $|\psi|^2$ can be interpreted as the probability density. Similar to wave-particle duality, students often describe the wave function as a sinusoidal particle path [41]. Table III presents reported misconceptions, divided into the two categories observed by Singh *et al.* [42] and Singh [43]: (1) misunderstanding due to overgeneralizations of prior concepts, and (2) difficulty distinguishing between closely related concepts [40–43], which results in a mix up of energy, wave functions, and probability. The first category corresponds with the work by Brooks and Etkina [36], who concluded classical metaphors cause misconceptions and promote misplaced classical thinking. This over-literal interpretation of classical metaphors was also observed by McKagan *et al.* [38]. These authors noticed that many students were likely to have difficulties in understanding the meaning of potential well graphs, and saw potential wells as external objects. McKagan *et al.* also observed that students mixed up wave functions and energy levels. Domert *et al.* [40] ascribed this to the use of diagrams combining energy levels and wave functions as illustrated in Fig. 1. However, McKagan *et al.* showed that eliminating these diagrams does not automatically prevent misconceptions.

TABLE III. Misconceptions about wave functions and potentials, categorized into two categories.

| | Overgeneralization of prior concepts | Mix-up of related concepts |
|---|---|---|
| Wave functions and potentials | Wave functions describe a trajectory [35,41] | Change in amplitude causes change in energy [38] |
| | Potential wells are objects [36,37] | The amplitude or equilibrium of the wave function is mixed up with energy [38] |
| | Height in potential graphs means position [35] | There is difficulty to distinguish between energy and probability [40] |
| Tunneling and probability | The amplitude of wave functions is a measure of energy [36,38,41] | Only the tops of the waves, which overtop the barrier, will pass [38,40] |
| | Probability is described with classical arguments (e.g., velocity) [35,40] | Part of the energy is reflected at a barrier during tunneling [38,40] |
| | Energy or effort is needed to tunnel through a barrier [37,38,44] | A single particle is described as an ensemble of particles [38,39] |

### 2. Tunneling and probability

Wave functions are not limited to classically permitted regions, they can extend past classical boundaries. This effect causes particles to have a probability of existing at positions that are classically impossible. An important result thereof is the phenomenon called tunneling; a small particle can end up on the other side of a classically impenetrable barrier. In this phenomenon no energy is lost and no work is done. In understanding of tunneling, the false belief that energy is lost during the process is prominent [37,38,44]. McKagan et al. [38] reported that students falsely attributed this energy loss to (1) work done on or by the particle inside the barrier; or to (2) the decrease of wave function amplitude. The same research also showed other misconceptions caused by a mix-up of physical quantities. Several students confused the wave function and energy. For example, some students erroneously believed that a decrease in amplitude causes an increase in energy, or the energy was partly reflected by the barrier. McKagan et al. also observed difficulty in understanding plane waves, which led to a mix-up of ensemble and single particle description. Domert et al. [40] observed that some students believed that only the tops of the waves, which supposedly were higher than the barrier, could pass the barrier. They also stated that misunderstanding of probability is an obstacle to the appropriate understanding of scattering and tunneling. They reported that many students had difficulty distinguishing between energy and probability, which they attributed in part to diagrams which mix wave functions and energy levels (see Fig. 1). Bao and Redish [35] and Wittmann et al. [39] observed that students can have difficulty with the predictability and stochastic nature of probability. Students falsely believed that the preceding distribution of outcomes influenced the subsequent outcome of single events, and tended to use classical arguments in their reasoning. This tendency was attributed to the lack of experience students have with probabilistic interpretations in physical systems.

### C. Atoms

The following section describes students learning difficulties related to the understanding of atomic structure, quantization, and spin, as found in the reviewed articles [12,24,25,31,45–56].

### 1. Atomic structure and models

The quantum atomic model describes the probability of observing the electron at a certain position, but it does not describe a temporal trajectory of an electron inside the atom. Research shows that secondary and undergraduate students hold on to various atom models [12,24,25,31,45–55] and can develop hybrid models consisting of combinations of different models [45]. Papageorgiou et al. [56] reported that the use of these models is influenced by the context of the task. The context of the question or previous questions influenced students' descriptions, which was also observed by McKagan *et al.* [48]. Based on a questionnaire administered to 140 undergraduate students, Ke *et al.* [46] divided the different atomic models into three different stages: (1) An early, planetary, quantum model, in which the electron orbits in a circle of constant radius, (2) a transitional model, in which the electron moves along a sinusoidal path, and (3) a probabilistic model, in which the position of the electron is uncertain. These stages are similar to the categories Ireson [24] observed. Additionally, Dangur *et al.* [54] divided the probabilistic model into a visual conceptual model based on probability distributions, and a mathematical model, in which students understand that the state of a particle can be described by a specific mathematical model. Although researchers used different classifications, one difficulty emerged in the majority of articles: Secondary and lower undergraduate students have difficulty letting go of Bohr's planetary atomic model [12,25,45–51,53,55]. Kalkanis *et al.* [12] ascribed this to many students believing that scientific content they learned previously is scientifically correct. This is in agreement with Stefani and Tsaparlis [50], who observed that models are sometimes seen as replicas of reality. Ke *et al.* [46] and Wang and Barrow [53] reported that more experienced students understood the difference between various models and could switch between them. McKagan *et al.* [48] claimed the solution is in comparing and contrasting different models, but also reported that students had difficulty understanding the reasons for the development of new atom models, which Taber [47] also reported in his research related to energy levels.

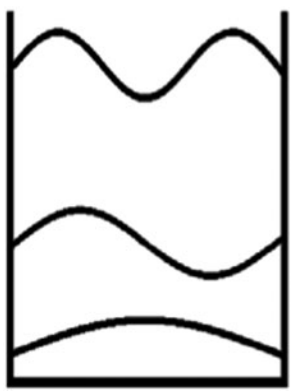

FIG. 1. A typical diagram as found in many textbooks, which simultaneously shows wave functions and energy levels.

#### 2. Energy levels, quantization and spin

To explain atomic spectra, current atomic models include energy levels. These energy levels cannot be arbitrary, but they have certain, specified values. These quantized energy levels can only be explained by considering them as bound wave functions and taking into account boundary conditions. Taber [47] observed that several secondary students did not understand the necessity of introducing quantization, because they did not see the planetary model as insufficient. Some students also had difficulty in forming an adequate concept of orbitals and confused orbitals with planetary orbits or concentric shells. Didiş *et al.* [55] reported that some undergraduate students did not understand that energy quantization is a natural phenomenon that occurs only when boundary conditions apply.

The distribution of electrons over the available energy levels in a system depends partly on electron spin. Spin is an intrinsic property of small particles and is a form of quantum angular momentum. But, in contrast to its classical counterpart, it is not a factual rotation. With regard to spin, Zhu and Singh [57], Taber [47], and Özcan [52] observed that many students falsely believed that quantum spin is an objects' rotation around its axis or around the core. Özcan indicated that there seemed to be a relation between the understanding of atomic models and spin. Those students who believed that quantum spin is an actual movement often used the classical atomic model. For students who described spin correctly, the use of the quantum atomic model was more dominant.

### D. Complex quantum behavior

The concepts discussed in the previous sections all are reductions from the fundamental principles of quantum mechanics. A wave function is a solution of the Schrödinger equation and represents a certain quantum state, which can be described by a set of quantum numbers. Little research has been done into misconceptions regarding these more complex subjects, such as quantum states, superposition and time evolution, for the secondary school level. Michelini *et al.* [58] developed and evaluated materials on quantum states and superposition, and concluded that secondary students' difficulties in accepting nondeterminism often cause a fall back to classical reasoning, and are an obstacle to understanding quantum states. Passante *et al.* [59] also researched understanding of quantum states and observed that undergraduate students find it hard to distinguish between pure superposition and mixed states. They also researched student understanding of time dependence, mainly focusing on upper division undergraduate level students [60]. One observation that could be useful for secondary and lower undergraduate education was that many students believed that for a time-dependent wave function, the probability of finding a particle in a region must also be time dependent. Regarding time dependence, Zhu and Singh [43,61] observed some students who falsely believed that after measurement the wave function will remain the same or, after collapsing, will eventually go back to its initial state.

## IV. RESEARCH TOOLS

This section answers the second subquestion: "What instruments have been designed and evaluated to probe student understanding on a conceptual level?" and presents an analysis of the questionnaires and instruments intended for secondary and lower undergraduate education that were observed in the 74 reviewed articles. The research tools are analyzed on how they are designed and evaluated, and on the topics which they cover. Table IV presents a summary of this analysis.

### A. Multiple-choice concept tests

Several concept tests have been designed and used to uncover students' difficulties, but a substantial part was only aimed at the upper undergraduate level and emphasized mathematical formalism [43,69–71]; other tests were not sufficiently evaluated [72]. The selected literature included three evaluated multiple choice questionnaires [2,16,68] suitable for secondary and lower undergraduate level students, which will be described in this section.

TABLE IV. Overview of research tools appropriate for probing conceptual understanding of secondary and lower undergraduate level students.

| Researchers | Year | Research tool | Level | Country | Content | Design and evaluation |
|---|---|---|---|---|---|---|
| Cataloglu and Robinet [2] | 2002 | QMVI | Undergraduate students | US | Wave functions, potential wells, quantization | Content based on existing materials and commonly used text books. Modified after student and faculty feedback and item analysis. Results suggested QMVI scores may be a reasonable measure of student understanding |
| Ireson [24,25] | 1999 | Multivariate analysis | Undergraduate students | UK | Wave-particle duality, atomic structure, quantization | Items based on previous research on students conceptions [62,63]. Multivariate analysis resulted in a holistic picture. Findings were consistent with other research, using different methodology. |
| McKagan et al. [16] | 2010 | QMCS | Undergraduate students | US | Wave-particle duality, wave functions, potential wells, atomic structure, quantization, measurement | Content based on literature, faculty interviews, textbook reviews and student observations. Modified after interviews, surveys and discussions. QMCS is too small to adequately probe student understanding. Useful as pretest and post-test for undergraduate students, but not for graduate students. |
| Sen [31] | 2002 | Concept map strategy | Undergraduate students | Turkey | Wave-particle duality, atomic structure | Strategy based on Ausubel's theory on cognitive and meaningful learning [64,65]. Reliability and validity were analyzed using Crohnbach's $\alpha$ and factor analysis. Results were consistent with another, questionnaire-based, study. |
| Taber [47] | 2005 | Typology of learning impediments | Upper secondary students | UK | Atomic structure | Typology based on consideration of the influence of prior knowledge [66]. Proposed modification: include substantive learning impediments categorized as analogical, epistemological, linguistic, pedagogical, or ontological. |
| Tsaparlis and Papaphotis [51] | 2009 | Questionnaire | Upper secondary students | Greece | Atomic structure | Content based on questions in an earlier study [67], which were judged for content validity by chemistry teachers. |
| Wuttiprom et al. [68] | 2009 | QPCS | Undergraduate students | Australia | Wave-particle duality | Content based on expert opinions and students difficulties Modified after trials with students and experts. Reliability was analyzed with item analysis, KR21 reliability test, and Ferguson's delta. |

### 1. Quantum Mechanics Visualization Inventory

Cataloglu and Robinett [2] designed the Quantum Mechanics Visualization Inventory (QMVI), based on existing materials and commonly used text books. Alterations to the preliminary inventory were made based on student feedback, comments from faculty colleagues and an item analysis. The QMVI consists of 25 questions and focuses on the interpretation of various diagrams. Although many of the questions require mathematical reasoning, approximately one-third of the questions address conceptual understanding of the influence of the potential energy on probability and the wave function. These questions can provide useful information on the student difficulties discussed in Sec. III B. The test was validated for content by content experts and Ph.D. candidates and analyzed for reliability and item difficulty in two pilot studies. The test was found to be reliable, but slightly difficult ($\alpha = 0.83$, mean item difficulty $= 0.45$). Afterwards, the QMVI was administered to students ranging from the sophomore level to the graduate level. Analysis showed there was a large correlation between the students' confidence in, and correctness of, their answers. Analysis also showed differences in understanding for the three different levels of instruction, which matched expectations. No articles were published on the evaluation of the QMVI at the secondary school level.

### 2. Quantum Mechanics Conceptual Survey

The Quantum Mechanics Conceptual Survey (QMCS) was designed to elicit student difficulties on topics covered in most courses on quantum mechanics [16]. For the preliminary version, textbooks were reviewed, students were observed, and faculty interviews were held to determine the topics. This preliminary version addressed wave functions, probability, wave-particle duality, the Schrödinger equation, quantization of states, the uncertainty principle, superposition, operators and observables, tunneling, and measurement. Over a period of three years this 25-item survey was altered, surveys were analyzed, and interviews were held with students. Finally, 12 questions proved to be useful for detecting student difficulties. The final questionnaire addresses the conceptual understanding of a broad range of topics discussed in Sec. III, i.e., wave-particle duality, wave functions, potential wells, atom structure, and quantization. Because of the small number of questions, however, the QMCS is not appropriate for proper statistical analysis and researchers suggested that more questions should be developed. The QMCS was tested at different levels, and the researchers concluded that the QMCS is a useful post-test for the upper undergraduate level. Preliminary results indicated it could also be suitable to investigate learning gains of lower undergraduate level students, but this needs to be verified in future research.

### 3. Quantum Physics Conceptual Survey

Wuttiprom *et al.* [68] developed the Quantum Physics Conceptual Survey (QPCS) to test student understanding of basic concepts of quantum mechanics. The researchers studied syllabi and consulted experts in order to determine topics and create survey questions. The QPCS addresses conceptual understanding of the photoelectric effect, wave-particle duality, the de Broglie wavelength, double slit interference, and the uncertainty principle, of which student difficulties were discussed in Sec. III A. The questions were trialed with different groups of students and each version of the survey was critiqued by a group of discipline or teaching experts to establish validity. Subsequently, the final survey, consisting of 25 items, was administered to 312 lower undergraduate students at the University of Sydney. The results were statistically analyzed for item difficulty, discrimination of single items, discrimination of the entire test and the consistency among the questions. Analysis showed that two items were likely to be too difficult and three items too easy (item difficulty index $> 0.9$ or $< 0.3$), five items also turned out to be poor discriminators (item point biserial coefficient $< 0.2$). Still, the KR-21 reliability index and Ferguson's delta were found to be satisfactory (KR21 $= 0.97$, $\delta = 0.97$). The researchers concluded that even though several items needed improvement, these results indicated that the QPCS is a reliable survey.

## B. Other tools

Besides multiple choice concept tests, there are other strategies to investigate students' difficulties. The reviewed literature included four other evaluated research tools, which emphasize students' reasoning, mental models, and underlying causes of misunderstanding [24,25,31,47,51].

### 1. Multivariate analysis

Ireson [24,25] designed a 40-item Likert-scale questionnaire, of which 29 items tested conceptual understanding of wave-particle duality, atom structure, and quantization. This questionnaire was administered to 338 lower undergraduate students. The analysis was based on the assumption that understanding can be represented by clustering the conceptions of a group of students. First, the responses were subjected to cluster analysis, which clusters individuals and gives insight into understanding at the group level. This resulted in three clusters, which were labeled quantum thinking, intermediate thinking, and mechanistic thinking. Second, Ireson used multidimensional scaling, which was used to map the response in multiple dimensions. This resulted in a two-dimensional model, of which the dimensions represented students' dual and nondeterministic thinking. This two-dimensional model confirmed the existence of three clusters; Ireson concluded that this method can be used to gain insight in students thinking and clusters or dimensions in their understanding.

#### 2. Concept map strategy

Sen [31] used a concept map strategy to evaluate the learning process, diagnose learning difficulties, and map the progression of students' cognitive structure. Training in creating concept maps was provided to 88 undergraduate students, from three different educational levels. At the end of the semester, the students each individually constructed a paper and pencil concept map. The concept map had to contain three main concepts (the atom, electron, and photon) and students were instructed to pay attention to the hierarchical order and links among concepts. Sen scored the concept maps for the number of valid concepts, relationships, branching, hierarchies, and cross-links. The scoring of the concept maps was tested for reliability, Cronbach's $\alpha$ was 0.67. Additionally, the scoring scheme was analyzed for construct validity by factor analysis. This analysis showed that the five scoring categories were correlated to separate single factors. The researcher also observed that the concept maps resembled results from a questionnaire-based study on the same subject. Results showed significant differences in the number of concepts and branches for the three different educational levels. Sen concluded that the results suggest that concept mapping can be used to investigate cognitive structures and the development thereof. However, the interpretation of the scores needs to be evaluated empirically [73].

#### 3. Typology of learning impediments

Taber [47] constructed and evaluated a typology of learning impediments, which he used to analyze underlying causes for students' difficulties. The typology was based on the Ausubelian idea that, for meaningful learning, students need to relate new concepts to prior knowledge. Four types of learning impediments were defined: (1) Students lack prerequisite knowledge; (2) students fail to make required connections; (3) students interpret the material inappropriately, because of their intuitive ideas; and (4) students interpret the material inappropriately, because of their cognitive structures. Taber used this typology to analyze data from an interview-based study on the understanding of chemical bonding of pre-university students. The researcher identified all four types of learning impediments and concluded that the typology is a useful heuristic tool, which can be used to interpret data on student learning. Still, Taber also recommended a refinement that takes into account misconceptions based on analogies or epistemological assumptions.

#### 4. Questionnaire on atomic structure

Tsaparlis and Papaphotis [51] designed a questionnaire for a study into the deep understanding and critical thinking of first-year undergraduates with regard to the quantum atom model. The questionnaire was based on a preliminary questionnaire that had been validated for content by chemistry teachers in a previous study [67]. It consisted of 14 open-ended questions; 9 of them were designed to test conceptual understanding, and the other questions were aimed at algorithmic knowledge. The questionnaire was administered to 125 students as part of a qualitative study. The researchers only drew conclusions about student understanding, the questionnaire itself was not evaluated.

### V. TEACHING STRATEGIES

This section addresses the subquestion: "What teaching strategies aimed at the secondary and lower undergraduate level have been tested, implemented and evaluated for their influence on student understanding?" and presents approaches promoting the understanding of quantum mechanical concepts that have been investigated in the selected literature. The following section presents the teaching strategies found in the selected articles, divided into instructional and multimedia-based strategies. There are several other activities described in literature, e.g., the hands-on activities from Visual Quantum Mechanics [74], the Dutch approach using the particle in a box [8], and the approach starting with qubits [75], but this review only discusses strategies which were implemented and evaluated in an educational setting.

#### A. Instructional strategies

There are still many questions concerning the teaching of introductory quantum mechanics. The introduction using wave-particle duality, for example, is still under discussion. Several alternative ways to introduce quantum mechanics have been used [58,76,77], but these alternatives have not been properly evaluated and compared to the use of wave-particle duality. However, several articles did describe investigations into the influence of teaching methods on student understanding. This section describes implemented and evaluated instructional strategies that were found within the selected literature [12,22,36,48,49,54,76,78–89], organized into four groups.

##### 1. Focus on interpretation

Because of quantum mechanics' indeterminacy, many interpretations are possible. Today's quantum experts do not support one single interpretation, although the Copenhagen interpretation is often considered to be the standard interpretation [90]. Baily and Finkelstein [78,79] researched the influence of addressing interpretations of quantum mechanics on student interpretations. Results showed that undergraduate students tended to prefer a local and deterministic interpretation if there was no emphasis on ontology. Baily and Finkelstein also presented results of the implementation of a new curriculum [76], which addressed the topic of "physical interpretation" explicitly. This curriculum included in-class discussions and experimental evidence, and aimed for understanding of different perspectives, their advantages, and limitations. Results of the use of this curriculum showed a clear change in student interpretation and the researchers concluded this confirms the importance of emphasis on interpretation. Greca and Freire [22] also researched the influence of teaching on undergraduate students' interpretations. For this purpose an interpretation was chosen that suited their didactic strategy, which emphasized a phenomenological-conceptual approach. The researchers used a realistic interpretation of the Copenhagen interpretation, in which the probability density function does not predict the probability of finding a particle, but the probability of the particle being present at a certain position. Comparison with a control group showed that in the experimental groups more students developed reasonable understanding. These examples showed the importance of an emphasis on interpretation in the design of new curricula.

### 2. Focus on models

Research showed that students tend to hold on to Bohr's planetary description of the atom [45,46,51,53], because it corresponds to students' classical worldview. Several approaches were evaluated to address this problem. Kalkanis *et al.* [12] presented an approach that emphasized the differences between classical and quantum mechanics. An instructional module focusing on the hydrogen atom was developed, which contrasted the classical and quantum models, and used the Heisenberg uncertainty relation as the basic principle. The module was taught to 98 preservice teachers and evaluated with pretests and post-tests and semistructured interviews. Results showed that a vast majority described the hydrogen atom correctly and could appropriately apply Heisenberg's uncertainty principle. The students had also become more aware of the process of learning and showed a change in worldview.

Strategies based on the historical development of the atomic model were evaluated by Unver and Arabacioglu [88] and McKagan *et al.* [48]. Unver and Arabacioglu developed a teaching module focusing on observations and experiments that led to alterations of the atomic model. The module was implemented in a course for preservice teachers ($N = 73$). Pretests and post-test comparisons showed a significant change in understanding. McKagan *et al.* designed an undergraduate course focusing on model building and reasoning for each model. Results showed that emphasis on the analysis of the predictions of each model, and the explanation of reasoning behind the development of the model, resulted in an increase in the use of the Schrödinger model.

Classical analogies are also used to promote understanding of the quantum atom model. Budde *et al.* [80] developed the Bremen teaching approach for upper secondary schools, which is based on similarities between the quantum atom model and liquids. Nine students were taught that atoms consist of electronium, a liquid substance, to promote the idea that an atom has a continuous nature, in which electrons are not moving. Budde *et al.* observed that some students described electronium as having a particle nature, but students still developed the conception that electrons are not moving. The researchers concluded that its focus on plausible aspects lead to high acceptance of the electronium model.

### 3. Focus on mathematical or conceptual understanding

Lower undergraduate and secondary students do not have extensive mathematical skills, which are an important part of quantum physics. This raises the question to what extent mathematical skills are needed for good understanding of quantum concepts. Studies have been done into the relation between mathematical and conceptual understanding of quantum concepts. Koopman *et al.* [84] observed that undergraduate students in a Quantum Chemistry course lacked mathematical skills, and they designed a remedial program. This program consisted of a diagnostic test, a prelecture, and online mathematics assignments. Students' results were monitored and commented upon. Students could consult a tutor and, if needed, additional explanation was scheduled. Koopman *et al.* observed a positive correlation between students' scores on the math assignments and the final exams ($N = 29$). From a comparison with student's grades for calculus, the researchers concluded that mathematical skills are necessary, but not sufficient for conceptual understanding. Papaphotis and Tsaparlis [49,86] researched the relation between algorithmic and conceptual understanding in high school chemistry. The study was conducted on 125 science students at the start of their first year at university. Students completed a questionnaire that addressed procedural knowledge and conceptual understanding. No correlation was found between their levels of procedural and conceptual performance. To investigate the effect of a nonmathematical approach on student understanding of the atomic structure, Dangur, Avargil, Peskin, and Dori [54,82] developed a teaching module focusing on real-life applications and visualization. This module was used for 122 secondary students and 65 undergraduate students. Results showed a significant improvement of understanding for both secondary and undergraduate students. Comparison with mathematically oriented undergraduates showed that the undergraduate test-group scored significantly higher on textual and visual understanding. This research suggests a conceptual, nonmathematical approach for teaching quantum mechanics can lead to adequate understanding.

#### 4. Use of activities

Active learning has become increasingly important in research into student engagement and understanding [91]. As a consequence, several reviewed articles described investigations into the influence of student activities on conceptual understanding. One example of active learning is the use of peer interaction. Shi [87] researched the influence of peer interaction on student understanding of duality and atomic models. Peer interaction was used once or twice a week during an undergraduate course on quantum mechanics. Students in the experimental group scored significantly higher than the control group on the post-test. Deslauriers and Wieman [81] investigated the effect of two different teaching methods on students' learning. One group ($N = 57$) was taught traditionally, while the other ($N = 67$) experienced interactive engagement methods (quizzes, simulations, clicker questions). The QMCS was used to test understanding, and comparison of the results for the two groups showed that the use of interactive engagement methods resulted in significantly higher scores. Yildiz and Büyükkasap [89] researched the influence of writing on understanding of the photoelectric effect. Pre-service teachers ($N = 36$) had to write a letter to senior high school students in which they explained the photoelectric effect. Results showed that these students scored significantly better on the post-test and exams than the control group. Gunel [83] explored differences in learning gains for two different writing tasks on Bohr's atomic model and the photoelectric effect ($N = 132$). The study indicated that secondary students who created a PowerPoint presentation had significantly higher learning gains than those who completed a summary report. Muller et al. [85] explored how well undergraduate students ($N = 40$) could learn from watching a video of a student-tutor dialogue on quantum tunneling. Results were compared to students who watched a traditional explanation. The students who watched the dialogue performed significantly better on the post-test. These results all suggest that active learning can contribute to better understanding of quantum concepts.

### B. Multimedia

Numerous multimedia applications have been designed for teaching quantum mechanics, but not all have been thoroughly evaluated. An overview of useful multimedia for quantum mechanics education was provided by Mason et al. [92]. The following section discusses evaluated multimedia found in the reviewed articles [5,27,32,33,38,57,58,77,93–100]. First PhET, QuILT, and QuVis are treated, which are databases covering a large number of topics. Then other separate simulations and teaching sequences using simulations will be discussed.

#### 1. PhET

McKagan et al. [98] described 18 simulations on fundamental principles, historical experiments, or applications of quantum mechanics developed in the PhET (Physics Education Technology) project. Most of them were developed for use in an undergraduate level course. These simulations were developed based on previous research, student interviews, and classroom testing. The interviews and classroom testing mainly focused on finding problems in the simulations, but some results of interviews and exams showed that several simulations ("Davisson-Germer: Electron Diffraction" and "Photoelectric Effect") resulted in better understanding. The researchers also noted that student interviews on the simulation "Quantum Tunneling and Wave Packets" suggested that guided activities could improve students' learning path when using the simulations. However, more research could still be done into the learning gains seen with the use of these simulations. The simulations on the photoelectric effect and tunneling were described more extensively. The simulation "Photoelectric Effect" was used for curriculum improvement [27]. This curriculum, based on active engagement techniques, resulted in better understanding of the photoelectric effect. However, students had difficulty linking this experiment to the particle behavior of light. The simulation "Quantum Tunneling and Wave Packets" was also part of an improved curriculum [38] that led to greater insight into students' difficulties on tunneling.

#### 2. QuILTs

Singh [32] described the development of QuILT's (Quantum Interactive Learning Tutorials) covering a broad range of subtopics. These tutorials, which were developed for undergraduate courses, consist of a combination of tasks, homework, Java applets, and pretests and post-tests. QuILTs were designed based on knowledge of student difficulties, and evaluated using pretests, post-tests, and student interviews. The multimedia applications used in the QuILT's were adapted from different sources (e.g., PhET [98] and Physlets [101]). Results of the pre-experimental evaluation of QuILTs on time development, the uncertainty principle, and the Mach-Zehnder interferometer showed a substantial change in performance. Zhu and Singh also evaluated a QuILT regarding the Stern-Gerlach experiment [57] and quantum measurement [100]. Both resulted in distinct improvement of understanding. Comparison of the results for students who went through the tutorial on quantum measurement with those for a control group showed that the QuILT resulted in better scores on the post-test.

#### 3. QuVis

Kohnle et al. [96,97] reported on the development of QuVis, which is a collection of interactive animations and visualizations for undergraduate students. Student interviews and observation sessions were used to optimize the interface design. Subsequently, the researchers investigated the influence of two simulations (the potential step and the finite well) on student understanding in a quasiexperimental setting. Two groups of students completed a diagnostic test: an experimental group, which worked with the animations, and a control group. Statistical analysis of the test results showed a significant relation between having worked with the simulations and performance on questions covering the corresponding subjects. In more recent work, Kohnle *et al.* [95] presented simulations regarding two-level quantum systems. They evaluated the learning gains resulting from use of a simulation on superposition states and mixed states. Results showed a substantial change in understanding.

#### 4. Simulations on atomic structure

Several simulations were designed to improve understanding of the atomic structure. Chen *et al.* [93] investigated the different effect of static and dynamic representations on understanding of atomic orbitals. The researchers compared two groups of secondary students. One group completed a learning activity using static 3D representations, while the second group worked with a dynamic 3D representation. Analysis of a pretest and post-test showed that both representations increased conceptual understanding. However, the researchers concluded that students who worked with the dynamic representations had more sophisticated mental models of the atom. Ochterski [99] used research-quality software (GaussView) and designed and evaluated two activities ($N = 95$, $N = 71$) to introduce orbitals and molecular shape to high school students. Pretests and post-tests for both activities showed an increase in understanding; Ochterski concluded that research-quality software can be effective, even if students have little background in chemistry.

#### 5. Teaching sequences using simulations

Other simulations were evaluated within the context of the design of a course. Malgieri *et al.* [77] described a teaching sequence using the Feynman sum over paths method. This sequence used simulations in GeoGebra, which included the photoelectric effect and the double-slit experiment. The eight-hour course was tested on preservice teachers ($N = 12$) and evaluated with a pretest and post-test. Results showed a good level of understanding of the role of measurement and the single photon interpretation of the double-slit experiment. However, the understanding of the uncertainty principle was still not adequate. Müller and Wiesner [5] designed and implemented a secondary school course using virtual experiments with the Mach-Zehnder interferometer and the double slit. Interviews and a questionnaire showed that students ($N = 523$) who took part in the course developed better quantum understanding than the control group. Michelini *et al.* [58] proposed a secondary school teaching sequence using prevision experiment comparison (PEC) strategies. This sequence included simulations on light interaction with Polaroids and Malus law. Analysis of student worksheets ($N = 300$) and a group discussion ($N = 17$) showed that the approach stimulated learning for at least 75% of the students. The researchers concluded that software simulations can help students in building a phenomenological framework, but are not sufficient.

#### 6. Quantum computer games

A different way of using multimedia is the use of quantum computer games. Gordon and Gordon [94] developed the computer game "Schrödinger cats and hounds" to teach quantum mechanical concepts in a fun way. Game-aided lectures were given to 95 undergraduate students. Analysis of a pretest and post-test showed an increase in understanding.

### VI. CONCLUSIONS

In this paper we presented an overview of existing knowledge on student difficulties, research tools for investigation of conceptual understanding, and teaching strategies. The conclusions of this literature review will be presented in this section.

#### A. Student difficulties

Analysis of the selected articles shows that secondary and undergraduate students have many difficulties when they learn quantum mechanics. Much research has been done into misunderstanding of wave-particle duality, wave functions, and atoms. However, not much research has been done into student difficulties with complex quantum behavior, and no research was found concerning secondary students' understanding of the wave function. Research into the understanding of wave-particle duality showed that undergraduate students' understanding can be clustered according to the extent of classical thinking [20,22,24–26]. Researchers also observed misplaced classical thinking in understanding of the wave function; several students displayed an over-literal interpretation of classical metaphors [36,38], or used classical reasoning in describing the process of tunneling [38,44]. Research into students' understanding of the quantum atomic model also indicated that both secondary and undergraduate students hold on to previously learned, semiclassical models [12,25,45–51, 53,55]. From these results we can conclude that many difficulties that students experience are related to the inability to connect quantum behavior to the physical reality as they see it, which results in a mix-up of classical and quantum concepts. Although this has been researched mainly for the undergraduate level, the existing research shows similarities in secondary and undergraduate students' understanding of duality and atomic models. This suggests that the mix up of classical and quantum concepts is also an important issue at the secondary level. Researchers have proposed several ideas concerning solutions for the mix up of classical and quantum concepts; e.g., analogies should be well defined [36], diagrams should be unambiguous [38,40], and students should have more knowledge of the use of models in physics [12,48,88]. However, the impact of these proposed solutions remains to be investigated.

TABLE V. Topics covered by the research tools.

|  |  | QMVI | QMCS | QPCS | Sen[a] | Ireson | Taber | Tsaparlis |
|---|---|---|---|---|---|---|---|---|
|  |  | Lower undergraduate education (●) | | | | Secondary education (■) | | |
| Wave-particle duality | Photons and electrons |  | ● | ● | ● | ●/■ | ■ |  |
|  | Double slit experiment |  | ● | ● | ● | ●/■ |  |  |
|  | Uncertainty principle |  | ● | ● | ● |  |  | ■ |
|  | Photoelectric effect |  |  | ● | ● |  |  |  |
| Wave functions | Wave functions and potential wells | ● | ● |  |  |  |  |  |
|  | Tunneling | ● | ● |  |  |  |  |  |
|  | Probability | ● | ● |  |  |  |  | ■ |
| Atoms | Atomic structure |  | ● |  | ● | ●/■ | ■ | ■ |
|  | Energy levels, quantization, and spin | ● | ● |  | ● | ●/■ | ■ | ■ |
| Complex QM behavior | Quantum states |  |  |  |  |  |  |  |
|  | Superposition |  |  |  |  |  |  |  |
|  | Time evolution and measurement | ● | ● |  |  |  |  |  |

[a]Dependent on individual student responses.

## B. Research tools

The research tools discussed in Sec. IV all include conceptual questions that could be useful probing the understanding of secondary and lower undergraduate level students. The topics addressed in these tools are wave-particle duality, wave functions, quantization, atomic structure, and measurement. Table V gives an overview of the topics covered by each research tool. As can be seen, none of the instruments covers the complete spectrum of quantum mechanics. Furthermore, only the research tools from Ireson, Taber, and Tsaparlis regarding duality and atomic structure, are used in secondary school settings. The QMVI addresses conceptual understanding only in part, and therefore some questions can be appropriate for the secondary and lower undergraduate level. The QMCS, which covers most of the topics, aims to probe conceptual understanding, but has not been thoroughly evaluated for secondary and lower undergraduate education. Moreover, the QMCS includes too few questions for statistical analysis. These results imply that the development and evaluation of more questions is needed, not only to cover all major topics from quantum mechanics, but also to make statistical analysis possible.

## C. Teaching strategies

Various methods and approaches have been designed and used to promote understanding in introductory courses on quantum mechanics, at both the secondary and undergraduate level. Still, only a small selection of these methods has been evaluated for their impact on students' understanding. These evaluations show the following:

(1) emphasis on interpretations influences undergraduate student perspectives, and should be taken into account in the development of curricula and teaching sequences;
(2) emphasis on the development of and the differences between various atomic models can result in better understanding of undergraduate students;
(3) a nonmathematical, conceptual approach can lead to adequate understanding for secondary and undergraduate students;
(4) active learning contributes to the understanding of quantum mechanical concepts.

However, there is a need for more empirical research into the teaching of quantum mechanics and teaching strategies should be researched for both secondary and undergraduate education.

Furthermore, many multimedia applications have been designed for teaching quantum mechanics. Table VI shows that for undergraduate education all quantum topics are covered by the multimedia applications found in the reviewed articles. For secondary education there are fewer applications and most topics are covered. Most of the applications were evaluated for practical use; only some of the simulations were also evaluated for their influence on student understanding. Singh and Zhu [32,57,100] have made a start with the design and evaluation of tutorials using multimedia, but more research into how these applications can be used to promote understanding is needed.

## D. Implications for researchers

This paper shows the current state of research into learning difficulties and teaching strategies for quantum physics at the secondary and lower university level. Analysis of 74 articles showed there are many groups researching student understanding, teaching strategies or assessment methods, mostly aiming at undergraduate education.

### 1. Lower undergraduate level

For lower undergraduate students, several learning difficulties were observed in the selected articles, but little research has been done into the conceptual understanding of complex quantum behavior. Although these topics are also difficult for upper-graduate students, it would be good to investigate to what extent these topics can be taught conceptually. More research should also be done into the underlying difficulties and causes of observed student difficulties. Several assessment methods have been designed for the undergraduate level, but there is still need for tests that cover more topics and are suitable for statistical analysis. More empirical research is needed for the further development of lower undergraduate level courses on quantum mechanics, in which teaching strategies are evaluated and compared using proper assessment tools. This research should also include investigations into ways to promote students' understanding using multimedia applications and experiments.

TABLE VI. Overview of quantum mechanical topics covered by the multimedia applications.

| | | PhET | QuILT[a] | QuVis | Malgieri | Gordon | Chen | Ochterski | Müller | Michelini |
|---|---|---|---|---|---|---|---|---|---|---|
| | | Lower undergraduate education (●) | | | | | Secondary education (■) | | | |
| Wave-particle duality | Photons and electrons | ● | ● | ● | ● | ● | | | ■ | ■ |
| | Double slit experiment | ● | ● | | ● | | | | ■ | |
| | Uncertainty principle | ● | ● | ● | ● | | | | ■ | ■ |
| | Photoelectric effect | ● | | ● | | | | | ■ | |
| Wave functions | Wave functions and potential wells | ● | ● | ● | | | | | ■ | |
| | Tunneling | ● | ● | ● | | | | | | |
| | Probability | ● | ● | ● | ● | | | | ■ | ■ |
| Atoms | Atomic structure | ● | ● | ● | | | ■ | ■ | ■ | |
| | Energy levels, quantization, and spin | ● | ● | ● | | | ■ | ■ | ■ | |
| Complex quantum behavior | Quantum states | ● | ● | ● | ● | ● | | | ■ | ■ |
| | Superposition | ● | ● | ● | ● | ● | | | ■ | ■ |
| | Time evolution and measurement | ● | ● | ● | | | | | | |

[a]Tutorials using simulations of other sources.

### 2. Secondary school level

With regard to quantum mechanics at the secondary school level, more empirical research into teaching strategies is also needed. But, although many learning difficulties that were found in research at the undergraduate level were confirmed for secondary school students, several topics have not yet been thoroughly investigated and more research into learning difficulties is needed. For the secondary school level, there is a need for more research into the understanding of wave functions and potential wells, topics that are part of several secondary school curricula. Research into the teaching of quantum states at a conceptual level is also needed, because this is part of some secondary school curricula.

To thoroughly investigate teaching strategies, multimedia applications, and experiments suitable for secondary school students, research tools are needed. The existing concept tests primarily focus on the undergraduate level, and therefore, it remains to be investigated whether these assessment tools are also applicable at the secondary school level.

### E. Implications for teachers

Analysis of the current research shows that students have many difficulties while learning quantum mechanics. Although most of the research has been conducted at the undergraduate level, overlapping research shows similar difficulties at both levels addressed in the studies reviewed. Therefore, both lower undergraduate and secondary school teachers can benefit from the research discussed here. This paper shows that there has been little empirical research into ways to promote understanding, but teachers should be aware that students tend to hold on to classical thinking, which leads to the misinterpretation of unfamiliar quantum concepts, and the mix up of classical and quantum physics. It can be helpful to emphasize differences and similarities between quantum concepts and students' preconceptions, which has proved to be useful in the teaching of the quantum atomic model at the undergraduate level. Teachers should also be aware that it is important to specify the limitations of metaphors, because they can lead to over-literal interpretations.

### ACKNOWLEDGMENTS

This work was funded by The Netherlands Organization for Scientific Research (NWO) under Grant No. 023.003.053.

### APPENDIX: OVERVIEW OF RESEARCH INTO STUDENT DIFFICULTIES`

See Table VII.

TABLE VII. Details of the selected articles on student difficulties described in Sec. III.

| Part | Researchers | Topic | Level | Country | Methodology and analysis |
|---|---|---|---|---|---|
| A | Asikainen and Hirvonen [19] | Photoelectric effect | Undergraduate students and physics teachers | Finland | A case study, using pretest and post-test and semistructured interviews, was carried out with preservice ($N = 8$) and in-service ($N = 17$) teachers. Test responses were categorized, interviews were used for validation. |
| | Ayene et al. [20] | Wave-particle duality, uncertainty principle | Undergraduate students | Ethiopia | Semistructured interviews were conducted with undergraduate students ($N = 25$). Responses were categorized. |
| | Dutt [21] | Wave-particle duality, double slit experiment, photoelectric effect, quantization | Upper secondary students | Australia | Test and worksheet data from grade 12 students were analyzed and interviews were held with 6 volunteering students. |
| | Greca and Freire [22] | Wave-particle duality, uncertainty principle, probability distribution, superposition | Undergraduate students | Brazil | Concept tests and conceptual problems were used ($N = 89$), field notes were collected during classes. Responses were categorized using hierarchical clustering and multidimensional scaling. |
| | Hubber [23] | Light | Upper secondary students | Australia | Three semistructured interviews conducted and two questionnaires were administered ($N = 6$). Responses were categorized. |
| A/C | Ireson [24] | Wave-particle duality, atoms | Undergraduate students | UK | A questionnaire was given to the students ($N = 338$). Responses were analyzed with cluster analysis and multidimensional scaling. |
| A/C | Ireson [25] | Wave-particle duality, atoms | Undergraduate students | UK | A questionnaire was given to the students ($N = 338$). Responses were analyzed using cluster analysis and multidimensional scaling. |
| | Johnston et al. [1] | Wave-particle duality | Undergraduate students | Australia | Students ($N = 33$) were given two short-response quizzes. Responses were categorized and analyzed for correctness. |
| | Mannila et al. [26] | Wave-particle duality | Undergraduate students | Finland | Intermediate level students ($N = 29$) answered 8 open-ended questions. Modified concept maps were created for each response, compared to a "master map" based on experts' conceptions and categorized. |
| | Masshadi and Woolnough [4] | Wave-particle duality | Upper secondary students | UK | Students ($N = 83$) were given a semistructured questionnaire. Responses were categorized. |
| | McKagan et al. [27] | Photoelectric effect | Undergraduate student | USA | After a reformed course, students' responses to two exam questions were analyzed ($N = 465, N = 188$). |
| | McKagan et al. [16] | Wave-particle duality, double slit experiment | Undergraduate students | USA | Interviews were conducted ($N = 46$) during the design and evaluation of the QMCS. |
| | Müller and Wiesner [5] | Wave-particle duality, atoms, uncertainty principle, non-determinism | Secondary and undergraduate students | Germany | A questionnaire was administered to secondary students ($N = 523$) and interviews were conducted with secondary students ($N = 27$) and undergraduates ($N = 37$). Responses were categorized. |
| | Oh [28] | Photoelectric effect | Undergraduate students | South Korea | Three groups of students ($N = 31$, $N = 49$, $N = 49$) were given a pretest and a post-test, which were validated by interviews. Responses were categorized. |

(Table continued)

TABLE VII. (Continued)

| Part | Researchers | Topic | Level | Country | Methodology and analysis |
|---|---|---|---|---|---|
| | Olsen [29] | Wave-particle duality | Upper secondary students | Norway | Students from 20 different schools ($N = 236$) were given a test. Multiple choice questions were analyzed quantitatively, open-ended questions were categorized. |
| | Özcan [30] | Photoelectric effect, blackbody radiation, Compton effect | Undergraduate students | Turkey | Preservice physics teachers ($N = 110$) were given a questionnaire. Responses were categorized and analyzed for correctness. |
| A/C | Sen [31] | Wave-particle duality, atoms | Undergraduate students | Turkey | Students ($N = 88$) created a concept map. These maps were analyzed for number of concepts, relationships, branches, hierarchies, and cross-links. |
| | Singh [32] | Uncertainty principle, time development, Mach-Zehnder interferometer | Undergraduate students | USA | A pretest and post-test were given to students ($N = 12$) who did the QuILT. Examples of students' responses were provided. |
| | Sokolowski [33] | Photoelectric effect | Upper secondary school | USA | A group of students ($N = 15$) answered one conceptual question during an assignment. Examples of responses were provided. |
| | Vokos et al. [34] | Double slit experiment | Undergraduate students | USA | Written problems were given to students ($N = 450$) in various physics undergraduate courses and interviews were conducted ($N = 14$). Students' reasoning was analyzed and categorized. |
| B | Bao and Redish [35] | Probability | Undergraduate students | USA | Interviews were conducted with physics students ($N = 16$). The observations were summarized. |
| | Brookes and Etkina [36] | Potential wells | Undergraduate students | USA | Students ($N = 4$) were observed while working on homework problems. Examples of students' reasoning were shown and analyzed. |
| | Domert et al. [40] | Probability, tunneling | Undergraduate students | Sweden | Students ($N = 12$) were interviewed while working with a computer simulation. Observations were categorized and examples were given. |
| | McKagan et al. [38] | Tunneling | Undergraduate students | USA | Data was collected for eight courses, consisting of observations, responses to essay questions, interviews, and a concept test (QMCS). Observations were categorized and illustrated, test results were reported. |
| | Özcan [41] | Wave functions, operators | Undergraduate students | Turkey | Semistructured interviews were held with preservice physics teachers ($N = 34$). Observations were categorized. |
| | Özcan et al. [37] | Potential wells | Undergraduate and graduate students | Turkey | A concept test was given to undergraduate ($N = 95$) and graduate ($N = 15$) students. Semi-structured interviews were held with 10 students. Student responses were presented. |
| B/D | Singh [43] | Wave functions, probability, measurement | Undergraduate and graduate student | USA | Surveys were administered to graduate students ($N = 202$), interviews were held with graduate and undergraduate students ($N = 15$). Results were categorized and examples were given. |
| | Singh et al. [42] | Wave functions, probability, measurement | Undergraduate and graduate students | USA | Surveys were administered to graduate ($N = 200$) and undergraduate ($N = 89$) students. Examples of difficulties were presented. |

(Table continued)

TABLE VII. (Continued)

| Part | Researchers | Topic | Level | Country | Methodology and analysis |
|---|---|---|---|---|---|
| | Wittmann et al. [39] | Probability | Undergraduate students | USA | Students ($N = 42$) were given a pretest and post-test. A series of questions were given also during the semester. Students' responses were presented. |
| | Wittmann et al. [44] | Tunneling | Undergraduate students | USA | Written examination questions, ungraded quizzes, surveys, and interviews were analyzed by content analysis, interpretation of diagrams, and descriptions of students' actions. |
| C | Dangur et al. [54] | Atomic structure | Upper secondary and undergraduate students | Israel | Pretest and post-test were used to probe secondary ($N = 122$) and undergraduate ($N = 65$) student understanding. A rubric was designed to analyze the 3-item test. |
| | Didiş et al. [55] | Light, energy, angular momentum | Undergraduate students | Turkey | Interviews were conducted, a test was administered and exams were analyzed ($N = 31$). The interviews were coded and mental models were constructed. |
| | Kalkanis et al. [12] | Atomic structure, models | Undergraduate students | Greece | A concept test was given to the test group ($N = 98$) and a control group ($N = 102$). Semistructured interviews were conducted with a sample of the test group. Difficulties found during the interviews were summarized. |
| | Ke et al. [46] | Atomic structure | Upper secondary—Ph.D. students | Taiwan | A questionnaire was given to students from high school to Ph.D. level ($N = 140$). Responses were categorized. Twenty-eight students were interviewed using concept cards in order to refine the categorization. |
| | McKagan et al. [48] | Atomic structure, models | Undergraduate students | USA | One exam question was analyzed for four courses ($N = 591$). Responses were categorized. |
| | Özcan [52] | Spin | Undergraduate students | Turkey | Interviews were conducted with introductory ($N = 24$) and advanced ($N = 25$) students. The results were categorized. |
| | Papageorgiou et al. [56] | Atomic structure | Upper secondary students | Greece | Students ($N = 421$) were given two cognitive tests measuring field dependence and reasoning abilities. A third test was used to assess students' representations of the atom. These representations were categorized and the influence of student characteristics thereon was investigated. |
| | Papaphotis and Tsaparlis [49] | Atomic structure, uncertainty principle | Undergraduate students | Greece | A questionnaire was given to first-year students ($N = 125$). Student difficulties were summarized and illustrated with examples. |
| | Petri and Niedderer [45] | Atom structure | Upper secondary students | Germany | Observations, questionnaires, interviews and written materials were analyzed to describe the learning pathway of one student within a course. The data were analyzed for change in conceptions and meta-cognitive beliefs. |
| | Papaphotis and Tsaparlis [50] | Atomic structure | Undergraduate students | Greece | Interviews were held with 2nd year students ($N = 19$). The responses were categorized. |
| | Taber [47] | Atomic structure | Upper secondary students | UK | Semistructured interviews were conducted with students ($N = 15$). A typology of learning impediments was used to categorize the response. |

(Table continued)

TABLE VII. (*Continued*)

| Part | Researchers | Topic | Level | Country | Methodology and analysis |
|---|---|---|---|---|---|
| | Tsaparlis and Papaphotis [51] | Atomic structure, uncertainty principle | Undergraduate students | Greece | A questionnaire was given to first-year students ($N = 125$). Semistructured interviews were conducted with a subsample ($N = 23$). Students' discussions were summarized and illustrated with examples. |
| | Wang and Barrow [53] | Atomic structure, chemical bonding | Undergraduate students | USA | Three diagnostic tests were used to analyze student understanding ($N = 159$). Interviews, using a think-aloud protocol and interviews about events, were conducted with a subsample ($N = 48$). Representations of conceptual frameworks were created and analyzed by axial coding. |
| | Zhu and Singh [57] | Spin, Stern-Gerlach experiment | Undergraduate and graduate students | USA | Surveys were administered ($n > 200$) and semistructured interviews were conducted with a subset of students. Results were used to design a tutorial. |
| D | Emigh *et al.* [60] | Time dependence | Undergraduate | USA | Four tasks were used to assess student understanding ($N_1 = 416$, $N_2 = 439$, $N_3 = 285$, $N_4 = 215$). The tasks were examined to identify difficulties, and these difficulties were categorized. |
| | Michelini *et al.* [58] | Quantum states, nonlocality | Upper secondary students | Italy | Students ($N = 17$) took part in group discussions of worksheets. Examples of student reasoning and a summary of the discussion were presented. |
| | Passante *et al.* [59] | Superposition | Undergraduate and graduate students | USA | A multiple choice question was used to explore the understanding of sophomores, juniors, and graduate students. Juniors ($N = 32$) were asked to consider four statements. Results for the multiple choice question and an overview of student reasoning regarding these statements were provided. |
| | Zhu and Singh [61] | Measurement | Undergraduate and graduate students | USA | Concept tests, quizzes, and tests were analyzed over several years. Interviews and informal discussions were conducted with a subset of students to investigate students' reasoning. An overview of the responses and students' reasoning is presented. |